\documentclass[uplatex]{article}

\usepackage[dvipdfmx]{graphicx}
\usepackage{wrapfig}
\graphicspath{{../Picture1/}}

\usepackage{multirow}
\usepackage{here}

\makeatletter
\newcommand{\figcaption}[1]{\def\@captype{figure}\caption{#1}}
\newcommand{\tblcaption}[1]{\def\@captype{table}\caption{#1}}
\makeatother

\usepackage{longtable}

\usepackage{enumitem}

\usepackage{color}

\makeatletter
\def\Hline{\noalign{\ifnum0=`}\fi\hrule\@height 3.\arrayrulewidth \futurelet\reserved@a\@xhline}
\makeatother

\newcommand{\ctext}[1]{\raise0.2ex\hbox{\textcircled{\scriptsize{#1}}}}

\usepackage[top=30truemm,bottom=30truemm,left=20truemm,right=20truemm]{geometry}
\usepackage{fancyhdr}
\usepackage{amsmath}
\usepackage{amssymb}
\usepackage{mathcomp}
\usepackage{mathtools}
\usepackage{cases}
\usepackage{cancel}
\usepackage{empheq}

\usepackage{comment}
\usepackage{authblk}

\author[1]{Shoki Iwaguchi}
\author[1]{Tomohiro Ishikawa}
\affil[1]{Department of Physics, Nagoya University, Nagoya, Aichi, 464-8602, Japan}
\author[2]{Masaki Ando}
\affil[2]{Department of Physics, The University of Tokyo, Bunkyo, Tokyo 113-0033, Japan}
\author[2]{Yuta Michimura}
\author[3]{Kentaro Komori}
\affil[3]{Institute of Space and Astronautical Science, Japan Aerospace Exploration Agency, Sagamihara, Kanagawa 252-5210, Japan}
\author[3]{Koji Nagano}
\author[4]{Tomotada Akutsu}
\affil[4]{Gravitational Wave Science Project, National Astronomical Observatory of Japan, Mitaka, Tokyo 181-8588, Japan}
\author[5]{Mitsuru Musha}
\affil[5]{Institute for Laser Science, The University of Electro-Communications, Chofu, Tokyo 182-8585, Japan}
\author[1]{Rika Yamada}
\author[1]{Izumi Watanabe}
\author[1]{Takeo Naito}
\author[1]{Taigen Morimoto}
\author[1]{Seiji Kawamura}

\title{Quantum noise in a Fabry-Perot interferometer
including the influence of diffraction loss of light}
\date{}
\begin{document}
\maketitle
\begin{abstract}
\noindent
The DECi-hertz Interferometer Gravitational wave Observatory (DECIGO) is designed to detect gravitational waves at frequencies between 0.1 and 10 Hz. In this frequency band, one of the most important science targets is the detection of primordial gravitational waves. DECIGO plans to use a space interferometer with optical cavities to increase its sensitivity. For evaluating its sensitivity, diffraction of the laser light has to be adequately considered. There are two kinds of diffraction loss: leakage loss outside the mirror and higher-order mode loss. These effects are treated differently inside and outside of the Fabry-Perot (FP) cavity. We estimated them under the conditions that the FP cavity has a relatively high finesse and the higher-order modes do not resonate. As a result, we found that the effects can be represented as a reduction of the effective finesse of the cavity with regard to quantum noise. This result is useful for optimization of the design of DECIGO. This method is also applicable to any FP cavities with a relatively small beam cut and the finesse sufficiently higher than 1.
\\
\\
\noindent
{\em keywords:}
Diffraction loss; Fabry-Perot cavity; quantum noise
\end{abstract}



\section{Introduction}

\ \ \ \ \ The DECi-hertz Interferometer Gravitational wave Observatory (DECIGO) is designed to detect gravitational waves at frequencies between 0.1 and 10 Hz. In this frequency band, one of the most important science targets is primordial gravitational waves \cite{Seto}. Observation of primordial gravitational waves is expected to provide crucial evidence for cosmic inflation theory. While observation of the primordial gravitational waves by ground-based detectors is challenging due to the ground vibration noise, pendulum thermal noise, etc., existing at low frequencies, and limited interferometer arm lengths, space interferometers enable observations by removing these obstacles. DECIGO also plans to use optical cavities between spacecraft to increase its sensitivity further.\\
\ \ \ \ \ The target sensitivity of DECIGO was established more than ten years ago to detect primordial gravitational waves. However, the recent observation of the cosmic microwave background (CMB) by the Planck satellite and other electromagnetic observations reduced the upper limit for primordial gravitational waves significantly \cite{Planck}  \cite{Kuroyanagi}. This reduction of the upper limit requires further improvement of the target sensitivity of DECIGO \cite{Yamada}. Therefore, we have been trying to improve the sensitivity by optimizing various parameters of DECIGO, such as the arm length, the laser power and the diameter, reflectivity, and mass of the mirrors.\\
\ \ \ \ \ For this optimization, we have to treat the diffraction loss of light in a Fabry-Perot (FP) cavity properly; we should treat the diffraction loss differently from other optical loss-related quantities such as absorption and transmission. For example, the part of the light that passes outside a mirror due to diffraction obviously does not cause radiation pressure noise. In contrast, light that is absorbed by the mirror causes radiation pressure noise. As for the light coming back to the input mirror from the end mirror of a FP cavity, the part of the light that misses the input mirror due to diffraction does not reach a photodetector positioned to sample light returning from the cavity. In contrast, return light that transmits through the input mirror is detected by the photodetector, and thus contributes to the shot noise at the photodetector. In previous investigations, the impact of the diffraction was not considered. However, it is important to consider the diffraction loss to more correctly design the sensitivity of an interferometer. For these reasons, it is essential to correctly calculate the quantum noise of an interferometer with the diffraction loss. The higher-order mode in a FP cavity is treated as loss in this paper on the condition that the beam cut by the diffraction in a FP cavity is small enough so that the finesse is sufficiently higher than 1.
The diffraction of the laser beam in the FP cavity is investigated in other paper such as \cite{diffaction}. In this paper, the treatment of the diffraction loss in the FP cavity in \cite{diffaction} is further developed for the calculation of quantum noise. In this paper, we will provide the proper treatment of diffraction loss in terms of quantum noise. It should be noted that this method is applicable to any FP cavities with a relatively small beam cut and the finesse sufficiently higher than 1. It should also be noted that noise sources other than quantum noise are not discussed here. In this paper, we discuss diffraction loss in a FP cavity at section 2, calculation of quantum noise including the effect of diffraction loss at section 3, result of quantum noise by using the DECIGO parameters in section 4, and summary of this paper in section 5.
\section{Treatment of diffraction loss in a FP cavity}

\ \ \ \ \ Diffraction influences the amplitude of laser light when the mirrors of FP cavity reflect or transmit it. Because in principle the laser beam extends to infinity in a plane perpendicular to the laser axis, the laser beam suffers a loss when transmitting through or reflecting from a mirror. Figure 1 illustrates a laser with the wavelength of $\lambda$ entering a FP cavity via the input mirror; it is a distance $l$ away from the beam waist of the cavity and has radius $R$ and the mirror has the amplitude transmissivity $t$. Assuming that the laser is a Gaussian beam with the Rayleigh length $z_\mathrm{R}$, the normalized absolute amplitude of the laser through the input mirror is given by

\begin{empheq}[left={\Psi (x,y,-l) =\empheqlbrace}]{alignat=2}
  t \sqrt{\frac{2 z_\mathrm{R}}{\lambda (l^2 + z_\mathrm{R}^2)}}  \mathrm{exp} \left[ - \frac{\pi z_\mathrm{R} \left( x^2 + y^2 \right)}{\lambda \left( l^2 + z_\mathrm{R}^2 \right)} \right] & \quad (x^2 + y^2 \leqq R^2)
    \label{eq:1} \\
  \sqrt{\frac{2 z_\mathrm{R}}{\lambda (l^2 + z_\mathrm{R}^2)}}  \mathrm{exp} \left[ - \frac{\pi z_\mathrm{R} \left( x^2 + y^2 \right)}{\lambda \left( l^2 + z_\mathrm{R}^2 \right)} \right] & \quad (x^2 + y^2 > R^2)
   \label{eq:2} \mathrm{.}
\end{empheq}

\begin{figure}[htbp]
  \centering
   \includegraphics[clip,width=12.0cm]{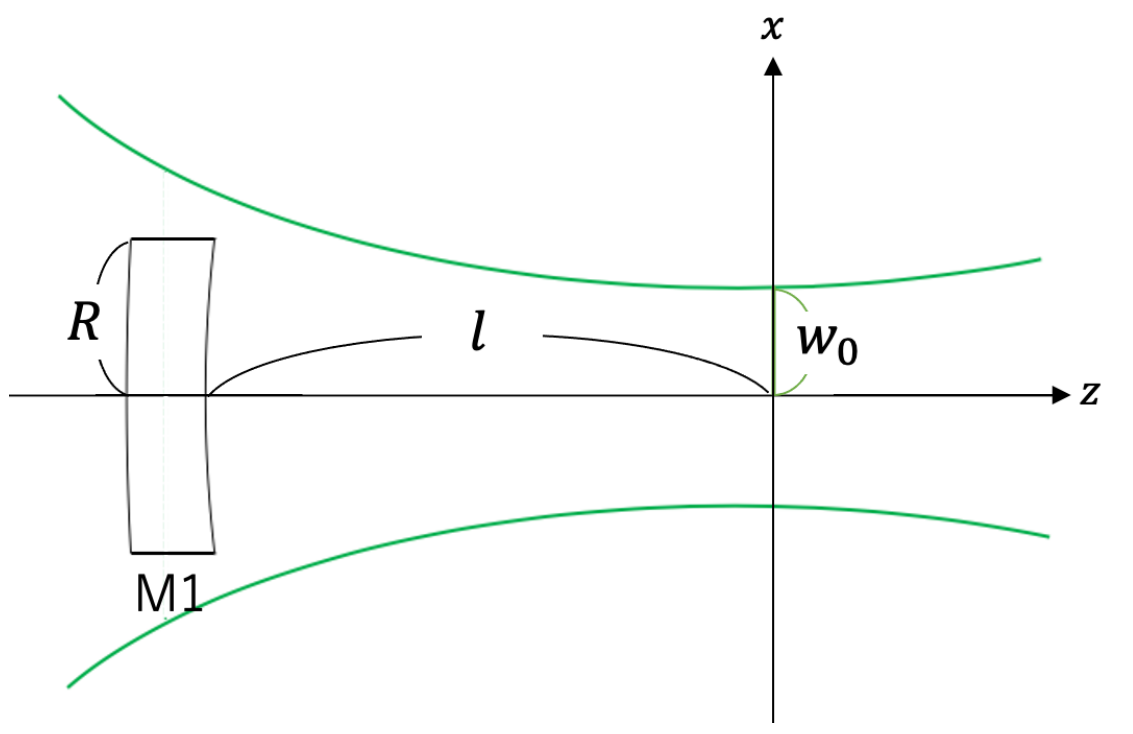}
   \caption{Illustration of the propagation of light with the wavelength of $\lambda$ around the input mirror of a FP cavity. The input mirror with radius of $R$ is labeled $\mathrm{M}1$. The $z$ axis is the laser beam axis, and the $xy$ plane is perpendicular to the laser beam axis. The beam waist in the cavity is at $z=0$.}
  \label{fig:1}
\end{figure}

When the resonant mode of the cavity is a $\mathrm{TEM}_{00}$ mode, we can treat only this fundamental mode under the condition that the FP cavity has relatively high finesse and the higher-order modes do not resonate in the FP cavity. The higher-order modes can be ignored in the FP cavity with high finesse because there is much more fundamental mode than the higher-order modes due to the substantial amplification of the $\mathrm{TEM}_{00}$ mode by the high finesse. Assuming that the contribution of the laser through the outside of the mirror is negligible, the normalized absolute amplitude of the fundamental mode is given by \cite{Laser}

\begin{equation}
    \Psi_{00} (x,y,z) = \sqrt{\frac{2 z_\mathrm{R}}{\lambda (z^2 + z_\mathrm{R}^2)}}  \mathrm{exp} \left[ - \frac{\pi z_\mathrm{R} \left( x^2 + y^2 \right)}{\lambda \left( z^2 + z_\mathrm{R}^2 \right)} \right]
      \label{eq:3} \mathrm{.}
\end{equation}

With Eq.(1) and Eq.(3), the amplitude of the fundamental mode after transmitting through the input mirror is given by

\begin{equation}
    \langle \Psi_{00} | \Psi \rangle = t \left( 1 - \mathrm{exp} \left[ - \frac{2 \pi z_\mathrm{R}}{\lambda \left( l^2 + {z_\mathrm{R}}^2 \right)} R^2 \right] \right)
     \label{eq:4} \mathrm{.}
\end{equation}

Therefore, the laser power after transmitting through the input mirror, $P$, is given by

\begin{equation}
    P = P_\mathrm{in}~t^2 \left( 1 - \mathrm{exp} \left[ - \frac{2 \pi z_\mathrm{R}}{\lambda \left( l^2 + {z_\mathrm{R}}^2 \right)} R^2 \right] \right)^2
     \label{eq:5} \mathrm{.}
\end{equation}

The effective transmissivity $t_\mathrm{eff}$, which is the transmissivity influenced by the diffraction, is given by

\begin{equation}
    t_\mathrm{eff} = t \left( 1 - \mathrm{exp} \left[ - \frac{2 \pi z_\mathrm{R}}{\lambda \left( l^2 + {z_\mathrm{R}}^2 \right)} R^2 \right] \right)
     \label{eq:6} \mathrm{.}
\end{equation}

\begin{figure}[htbp]
    \centering
    \includegraphics[clip,width=12cm]{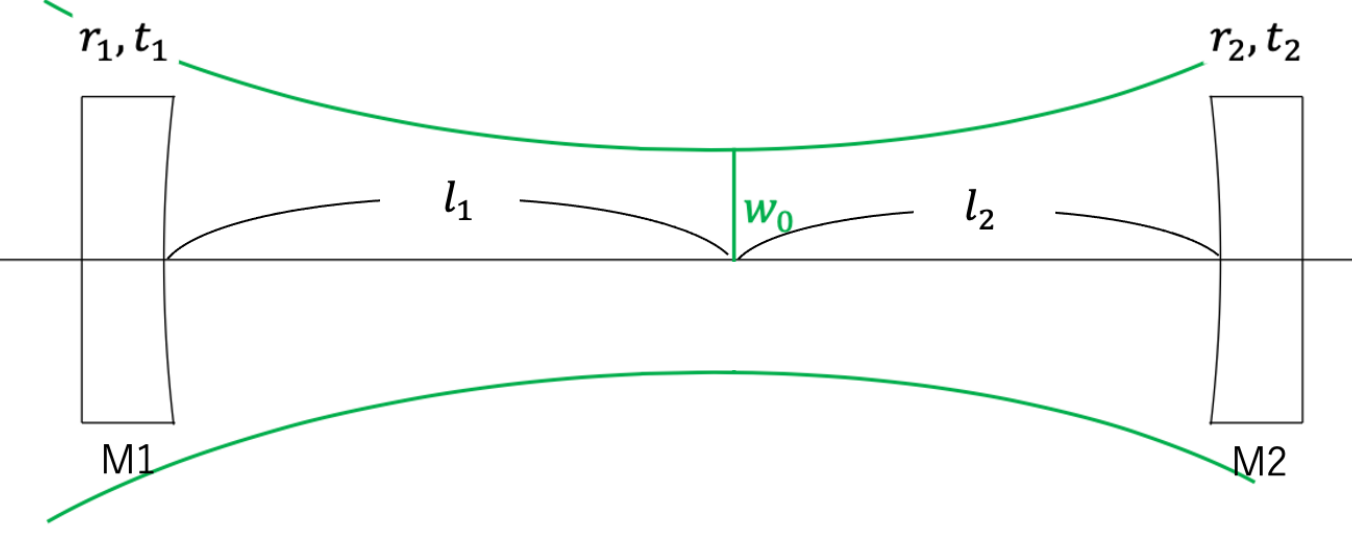}
    \label{fig:2}
    \caption{Configuration of the light around the FP cavity. The cavity has two mirrors. $\mathrm{M}1$ is the input mirror with reflectivity of $r_1$, transmissivity of $t_1$, and radius $R_1$. $\mathrm{M}2$ is the end mirror with reflectivity of $r_2$, transmissivity of $t_2$, and radius $R_2$. $w_0$ is the beam size at the beam waist. The distance from the mirrors to the beam waist are $l_1$ and $l_2$.}
\end{figure}


We define a diffraction loss factor, $D_\mathrm{i}$, for each mirror ($\mathrm{i}=1,2$) shown in Fig.2, given by

\begin{equation}
    D_\mathrm{i} \equiv \sqrt{1 - \mathrm{exp} \left[ - \frac{2 \pi z_\mathrm{R}}{\lambda \left( {l_\mathrm{i}}^2 + {z_\mathrm{R}}^2 \right)} R_\mathrm{i}^2 \right]}
  \label{eq:7} \mathrm{.}
\end{equation}

In the FP cavity, there are two kinds of effects of diffraction loss: leakage loss outside the mirror as expressed by Eq.(1) and higher-order mode loss as expressed by Eq.(3). The leakage loss has to be taken into account when the laser light is reflected or transmitted. Only inside the cavity, higher-order mode loss must be considered with leakage loss when the laser light is reflected by a mirror or transmits through a mirror. While we should treat only $D_\mathrm{i}$ as the leakage loss when we calculate the electric field outside the FP cavity, we have to treat ${D_\mathrm{i}}^2$ as the leakage loss coupled with the higher-order mode loss due to the cavity mode. The effective reflectivity $r_{\mathrm{eff},\mathrm{i}}$ and transmissivity $t_{\mathrm{eff},\mathrm{i}}$ shown schematically in Fig. 2, are defined by

\begin{align}
  &  r_{\mathrm{eff},\mathrm{i}} = r_\mathrm{i} \left( 1 - \mathrm{exp} \left[ - \frac{2 \pi z_\mathrm{R}}{\lambda \left( {l_\mathrm{i}}^2 + {z_\mathrm{R}}^2 \right)} R_\mathrm{i}^2 \right] \right) = r_\mathrm{i} {D_\mathrm{i}}^2   \label{eq:8} \ \  \mathrm{and} \\
  &  t_{\mathrm{eff},\mathrm{i}} = t_\mathrm{i} \left( 1 - \mathrm{exp} \left[ - \frac{2 \pi z_\mathrm{R}}{\lambda \left( {l_\mathrm{i}}^2 + {z_\mathrm{R}}^2 \right)} R_\mathrm{i}^2 \right] \right) = t_\mathrm{i} {D_\mathrm{i}}^2  \label{eq:9}  \mathrm{.}
\end{align}

\begin{figure}[htbp]
  \centering
   \includegraphics[clip,width=12cm]{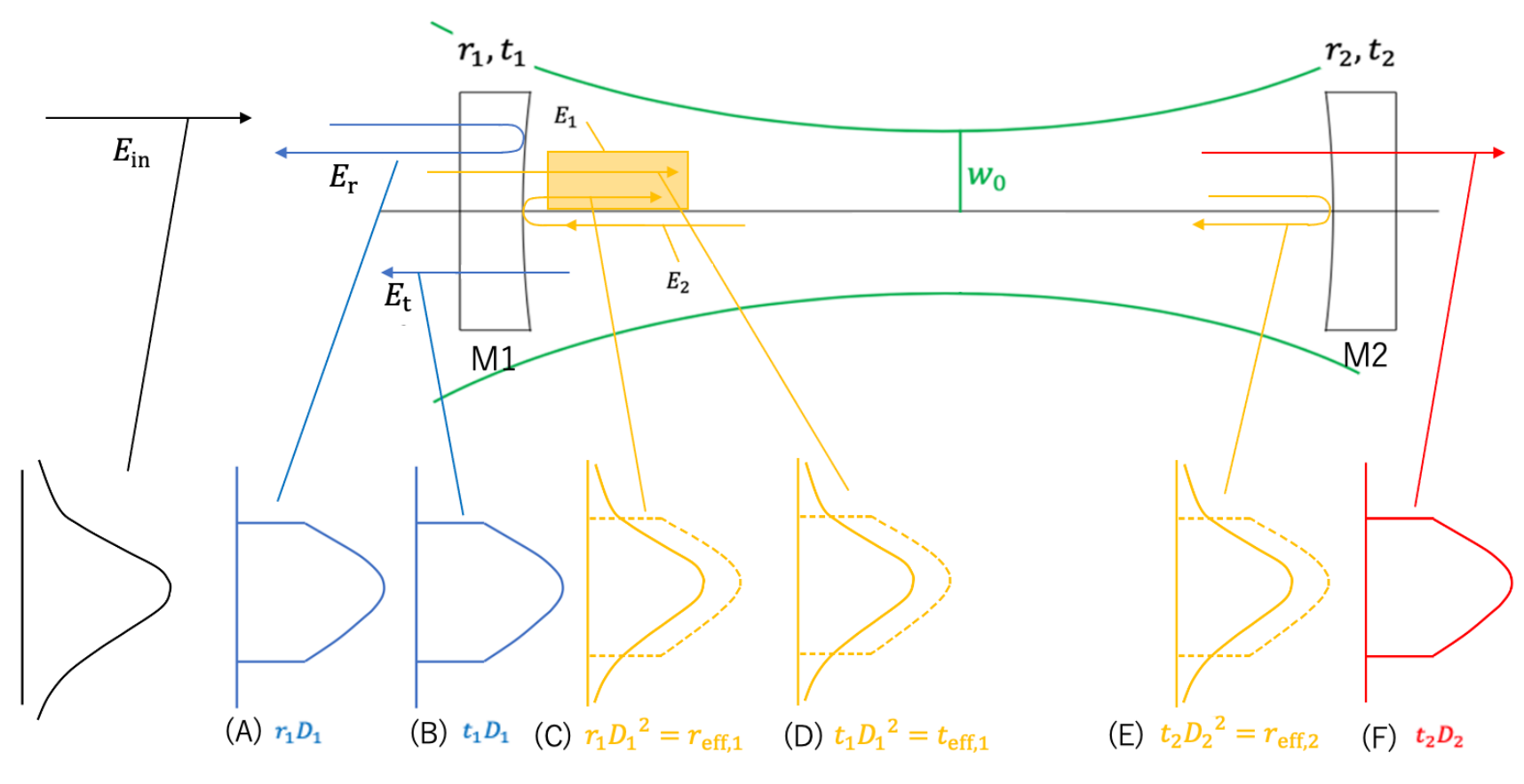}
   \label{fig:3}
   \caption{Illustration of each electric fields at each point around the FP cavity when the electric field $E_\mathrm{in}$ enters the input mirror ($\mathrm{M}1$) together with the transverse shape of the laser field at each point. $E_1$ is the electric field which is the sum of $E_\mathrm{in}$ through $\mathrm{M}1$ and goes back and forth $n$ times in the FP cavity. $E_2$ is $E_1$, which is reflected by $\mathrm{M}2$ and goes back and forth once. $E_\mathrm{t}$ is the fraction of $E_2$ transmitted by $\mathrm{M}1$. $E_\mathrm{r}$ is the fraction of $E_\mathrm{in}$ reflected by $\mathrm{M}1$. Locations outside the cavity, $\mathrm{(A)}$, $\mathrm{(B)}$, and $\mathrm{(F)}$, are affected by the leakage loss. Locations inside the cavity, $\mathrm{(C)}$, $\mathrm{(D)}$, and $\mathrm{(E)}$, are affected by both leakage loss and higher-order mode loss.}
\end{figure}


We calculate each electric field at each point of the cavity, as shown in Fig. 3, in preparation for the calculation of the quantum noise. With the round-trip phase change of the laser field defined as $\phi$, $E_1$ is given by

\begin{align}
    E_1 &= t_{\mathrm{eff},1} E_\mathrm{in} + t_{\mathrm{eff},1} E_\mathrm{in} \cdot \left( r_{\mathrm{eff},1} r_{\mathrm{eff},2} \cdot \mathrm{e}^{i \phi} \right) + \cdots \notag \\
        &= \frac{t_{\mathrm{eff},1}}{1 - r_{\mathrm{eff},1} r_{\mathrm{eff},2} \cdot \mathrm{e}^{i \phi}} E_\mathrm{in}
  \label{eq:10} \mathrm{.}
\end{align}

$E_2$, $E_\mathrm{t}$, and $E_\mathrm{r}$ are given by

\begin{equation}
    E_2 = E_1 \cdot r_{\mathrm{eff},2} \mathrm{e}^{i \phi} = \frac{t_{\mathrm{eff},1} r_{\mathrm{eff},2} \cdot \mathrm{e}^{i \phi}}{1 - r_{\mathrm{eff},1} r_{\mathrm{eff},2} \cdot \mathrm{e}^{i \phi}} E_\mathrm{in}
  \label{eq:11} \ \  \mathrm{,}
\end{equation}

\begin{equation}
    E_\mathrm{t} = E_2 \cdot \left(t_1 D_1\right) = \frac{\left( {t_1}^2 {D_1}^3 \right) r_{\mathrm{eff},2} \cdot \mathrm{e}^{i \phi}}{1 - r_{\mathrm{eff},1} r_{\mathrm{eff},2} \cdot \mathrm{e}^{i \phi}} E_\mathrm{in}
  \label{eq:12} \ \  \mathrm{and}
\end{equation}

\begin{equation}
    E_\mathrm{r} = E_\mathrm{in} \left(-r_1 D_1\right)
  \label{eq:13} \mathrm{.}
\end{equation}

$E_\mathrm{r}$ is multiplied by the negative reflectivity because of reflection from the High Reflection (HR) surface in this case coming from the higher index side instead of the condition where the reflection is coming from the air or vacuum side of the HR coating. In Eq.(12) and Eq.(13), we should treat only the leakage loss as the diffraction loss because the electric field detected at the photodetector includes the higher-order mode. For this reason, this electric field is multiplied by the coefficient $D_\mathrm{i}$ once. With Eqs.(12)-(13), the electric field of interference light $E_\mathrm{PD}$ is given by

\begin{align}
    E_\mathrm{PD} &= E_\mathrm{r} + E_\mathrm{t} \notag \\
        &= D_1 \left[ -r_1 + \frac{\left( {t_1}^2 {D_1}^2 \right) r_{\mathrm{eff},2} \cdot \mathrm{e}^{i \phi}}{1 - r_{\mathrm{eff},1} r_{\mathrm{eff},2} \cdot \mathrm{e}^{i \phi}} \right] E_\mathrm{in}
  \label{eq:14} \mathrm{.}
\end{align}

Then the power $P_\mathrm{PD}$ of the interference light is given by

\begin{align}
  P_\mathrm{PD} = {D_1}^2 \left[ {r_1}^2 - \left({t_1}^2 {D_1}^2\right) r_1 r_{\mathrm{eff},2} \left( \frac{\mathrm{e}^{-i\phi}}{1 - r_{\mathrm{eff},1} r_{\mathrm{eff},2} \cdot \mathrm{e}^{i \phi}} + \frac{\mathrm{e}^{i\phi}}{1 - r_{\mathrm{eff},1} r_{\mathrm{eff},2} \cdot \mathrm{e}^{-i\phi}} \right) \right. \notag \\
  \qquad\qquad\left. + \frac{\left\{ \left( {t_1}^2 {D_1}^2 \right) r_{\mathrm{eff},2} \right\}^2}{\left( 1 - r_{\mathrm{eff},1} r_{\mathrm{eff},2} \cdot \mathrm{e}^{i \phi} \right) \left( 1 - r_{\mathrm{eff},1} r_{\mathrm{eff},2} \cdot \mathrm{e}^{-i \phi} \right)} \right] P_\mathrm{in}
  \label{eq:15} \mathrm{.}
\end{align}

With the coefficient to simplify the formula, $\mathrm{F}$, defined as

\begin{equation}
    \mathrm{F} \equiv \frac{ 4 r_{\mathrm{eff},1} r_{\mathrm{eff},2} }{ (1 - r_{\mathrm{eff},1}r_{\mathrm{eff},2} )^2 }
  \label{eq:16}  \mathrm{,}
\end{equation}



Eq.(15) can be written as

\begin{align}
    P_\mathrm{PD}
    = \frac{\left[ r_1 - \left( {t_1}^2 + {r_1}^2 \right) {D_1}^2 \cdot r_{\mathrm{eff},2} \right]^2 + 4 r_{\mathrm{eff},1} r_{\mathrm{eff},2} \left( {t_1}^2 + {r_1}^2 \right) \sin^2 \left( \frac{\phi}{2} \right)}{\left( 1 - r_{\mathrm{eff},1} r_{\mathrm{eff},2} \right)^2 \left[ 1 + \mathrm{F} \sin^2\left(\frac{\phi}{2}\right) \right]} \cdot {D_1}^2 P_\mathrm{in}~.
  \label{eq:17}
\end{align}

Here we derive the finesse of the FP cavity, including the effect of diffraction loss. We define the effective finesse as $\mathcal{F}_\mathrm{eff}$. The finesse is $\nu_\mathrm{FSR} / \Delta \nu$ ; $\nu_\mathrm{FSR}$ is free spectral range and $\Delta \nu$ is the cavity bandwidth. First, we derive these terms, including diffraction loss. Using Eq.(10), the laser power $P_1$ inside the FP cavity can be written as

\begin{equation}
    P_1 = \frac{{t_{\mathrm{eff},1}}^2}{\left( 1 - r_{\mathrm{eff},1} r_{\mathrm{eff},2} \right)^2 \left[ 1 + \mathrm{F} \sin^2 \left( \frac{\phi}{2} \right) \right]} P_\mathrm{in}
  \label{eq:18} \mathrm{.}
\end{equation}

The length of FP cavity is $L=l_1 + l_2$. When the frequency of the laser is defined as $\nu$, the round-trip phase change $\phi$ is given by

\begin{equation}
   \phi (\nu) = - \frac{4 \pi L}{c} \nu
  \label{eq:19} \mathrm{.}
\end{equation}

The half of the maximum laser power is equal to a laser power derived from substituting Eq.(19) and $\nu = \Delta \nu / 2$ for Eq.(18), which is represented as

\begin{equation}
    \frac{1}{2} \cdot \frac{{t_{\mathrm{eff},1}}^2}{\left( 1 - r_{\mathrm{eff},1} r_{\mathrm{eff},2} \right)^2} P_\mathrm{in} = \frac{{t_{\mathrm{eff},1}}^2}{\left( 1 - r_{\mathrm{eff},1} r_{\mathrm{eff},2} \right)^2 \left[ 1 + \mathrm{F} \sin^2 \left(-\frac{2 \pi L}{c} \frac{\Delta \nu}{2} \right) \right]} P_\mathrm{in}
   \notag
\end{equation}
\begin{equation}
    \therefore~\left| \sin\left( - \frac{\pi L}{c} \Delta \nu \right) \right| = \frac{1}{\sqrt{\mathrm{F}}} .
  \label{eq:20}
\end{equation}

Assuming that $\Delta \nu \ll \nu_{\mathrm{FSR}}$, and using Eq.(20), the cavity bandwidth, $\Delta \nu$ is given by

\begin{equation}
    \Delta \nu \approx \frac{c}{\pi L} \cdot \frac{1}{\sqrt{\mathrm{F}}} = \frac{c}{2L} \cdot \frac{1 - r_{\mathrm{eff},1} r_{\mathrm{eff},2}}{\pi \sqrt{r_{\mathrm{eff},1} r_{\mathrm{eff},2}}} .
  \label{eq:21}
\end{equation}

And $\nu_\mathrm{FSR}$ is written as Eq.(22) by substituting Eq.(19) for $\phi(\nu) = \phi / 2$ and $\nu = \nu_\mathrm{FSR}$:

\begin{equation}
    \nu_\mathrm{FSR} = \frac{c}{2L}~.
  \label{eq:22}
\end{equation}

As a result, $\mathcal{F}_\mathrm{eff}$ is written as Eq.(23) with Eq.(21) and Eq.(22):

\begin{equation}
    \mathcal{F}_\mathrm{eff} \equiv \frac{\nu_\mathrm{FSR}}{\Delta \nu} = \frac{\pi \sqrt{r_{\mathrm{eff},1} r_{\mathrm{eff},2}}}{1 - r_{\mathrm{eff},1} r_{\mathrm{eff},2}} .
  \label{eq:23}
\end{equation}

This result shows that the effective finesse $\mathcal{F}_\mathrm{eff}$ is equal to the finesse $\mathcal{F}$ except for the difference between the reflectivity and the effective reflectivity.

\section{Quantum noise including diffraction loss}

\ \ \ \ We derive the frequency response to gravitational waves in FP interferometers as another preparation for the calculation of the quantum noise. The time $\Delta t_n$ is defined as the round trip time between the input and end mirrors, multiplied by $n$. When gravitational waves, $h(t)$, arrive at FP interferometers, the time, which takes for the laser light round trip, is given by

\begin{equation}
    \int_{t-\Delta t_n}^{t} \left( 1 - \frac{1}{2} h\left(t'\right) \right) \mathrm{d} t' \approx \frac{2 L n}{c}
  \label{eq:24} \mathrm{.}
\end{equation}

With Eq.(24), $\Delta t_n$ is given by \cite{Takaki} \cite{Ando}

\begin{align}
    \Delta t_n &= \frac{2 L n}{c} + \frac{1}{2} \int_{t-\Delta t_n}^{t} h\left(t'\right) \mathrm{d} t' \notag \\
        &\approx \frac{2 L n}{c} + \frac{1}{2} \int_{t-\frac{2 L n}{c}}^{t} h\left(t'\right) \mathrm{d} t' \notag \\
        &= \frac{2 L n}{c} + \int_{-\infty}^\infty \frac{1 - \mathrm{exp}\left[ -i \frac{2 L \omega}{c} n \right]}{i \cdot 2 \omega} h (\omega) \mathrm{d} \omega~.
  \label{eq:25}
\end{align}

The electric field of the interferometer light is given by a series like Eq.(10), with $ n \phi =- \Omega \Delta t_n$,

\begin{align}
    E_\mathrm{PD}
        &= -r_1 D_1 E_\mathrm{in} + t_1 D_1 \cdot t_{\mathrm{eff},1} \cdot r_{\mathrm{eff},2} \mathrm{e}^{i \phi} \cdot E_\mathrm{in} \sum_{k = 0} \left( r_{\mathrm{eff},1} r_{\mathrm{eff},2} \cdot \mathrm{e}^{i \phi} \right)^k \notag \\
        &= D_1 \left[ -r_1 + \left( {t_1}^2 {D_1}^2 \right) r_{\mathrm{eff},2} \sum_{n = 1} \left( r_{\mathrm{eff},1} r_{\mathrm{eff},2} \right)^{n-1} \mathrm{e}^{-i \Omega \Delta t_n} \right] E_\mathrm{in}
  \label{eq:26} \mathrm{.}
\end{align}

Using Eq.(25), this can be rewritten as

\begin{align}
    E_{\mathrm{PD}}
        &\approx D_1 \left[ -r_1 + \left( {t_1}^2 {D_1}^2 \right) r_{\mathrm{eff},2} \cdot \mathrm{e}^{-i \frac{2 L \Omega}{c}} \sum_{n = 1} \left( r_{\mathrm{eff},1} r_{\mathrm{eff},2} \cdot \mathrm{e}^{-i \frac{2 L \Omega}{c}} \right)^{n-1} \right. \notag \\
        &\qquad\qquad\qquad\left. \times \left\{ 1 - \int_{-\infty}^\infty \frac{\Omega}{2\omega} \left( 1 - \mathrm{exp} \left[ -i \frac{2L\omega}{c}n \right] \right) h(\omega) \mathrm{e}^{i\omega t} \mathrm{d} \omega \right\} \right] E_\mathrm{in} \notag \\
        &= D_1 \left[ -r_1 + \frac{\left( {t_1}^2 {D_1}^2 \right) r_{\mathrm{eff},2} \cdot \mathrm{e}^{-i \frac{2 L \Omega}{c}}}{1 - r_{\mathrm{eff},1} r_{\mathrm{eff},2} \cdot \mathrm{e}^{-i \frac{2 L \Omega}{c}}} + \left( {t_1}^2 {D_1}^2 \right) r_{\mathrm{eff},2} \cdot \mathrm{e}^{-i \frac{2 L \Omega}{c}} \int_{-\infty}^\infty \frac{\Omega}{2\omega} A h(\omega) \mathrm{e}^{i\omega t} \mathrm{d}\omega \right] E_\mathrm{in}
  \label{eq:27} \mathrm{.}
\end{align}

Here $A$, the coefficient to simplify the formula, is given by

\begin{align}
    A &= \sum_{n=1} \left\{ \left( r_{\mathrm{eff},1} r_{\mathrm{eff},2} \cdot \mathrm{e}^{-i \frac{2 L \Omega}{c}} \right)^{n-1} - \mathrm{e}^{-i \frac{2 L \omega}{c}} \left( r_{\mathrm{eff},1} r_{\mathrm{eff},2} \cdot \mathrm{e}^{-i \frac{2 L \left(\Omega + \omega \right)}{c}} \right)^{n-1} \right\} \notag \\
        &= \frac{1 - \mathrm{e}^{-i \frac{2L\omega}{c}}}{\left(1 - r_{\mathrm{eff},1} r_{\mathrm{eff},2} \cdot \mathrm{e}^{-i \frac{2 L \Omega}{c}}\right) \cdot \left( 1 - r_{\mathrm{eff},1} r_{\mathrm{eff},2} \cdot \mathrm{e}^{-i \frac{2 L \left(\Omega + \omega\right)}{c}} \right)}
  \label{eq:28} \mathrm{.}
\end{align}



Here $\alpha_\mathrm{C}$, the coefficient to simplify the formula, is given by

\begin{equation}
    \alpha_\mathrm{C} \equiv \frac{\left( {t_1}^2 {D_1}^2 \right) r_{\mathrm{eff},2}}{1 - r_{\mathrm{eff},1} r_{\mathrm{eff},2}}
  \label{eq:29} \mathrm{.}
\end{equation}

Also, we assume this condition given by

\begin{align}
    \frac{L \Omega}{c} = \mathrm{m} \pi~~(\text{m = integers})
  \label{eq:30} \mathrm{.}
\end{align}

Using Eq.(29) and (30), $E_\mathrm{PD}$ can be written as

\begin{align}
    E_\mathrm{PD}
    &= D_1 \left( -r_1 + \alpha_\mathrm{C} \right) \left[ 1 + i \cdot \frac{\alpha_\mathrm{C}}{-r_1 + \alpha_\mathrm{C}} \int_{-\infty}^\infty \frac{\Omega}{\omega} \frac{\sin\left( \frac{L\omega}{c} \right) \mathrm{e}^{-i \frac{L \omega}{c}}}{1 - r_{\mathrm{eff},1} r_{\mathrm{eff},2} \cdot \mathrm{e}^{-i \frac{2 L \omega}{c}}} h(\omega) \mathrm{e}^{i\omega t} \mathrm{d}\omega \right] E_\mathrm{in} \notag \\
        &\approx D_1 \left( -r_1 + \alpha_\mathrm{C} \right) \mathrm{exp} \left[ i \frac{\phi (t)}{2} \right] E_\mathrm{in}
  \label{eq:31} \mathrm{.}
\end{align}


Next we derive $h_\mathrm{shot} (f)$, which is the strain sensitivity of the shot noise in FP interferometer. First, we calculate $h_\mathrm{shot} (f)$ for one arm of a Fabry-Perot Michelson Interferometer (FPMI). Then we calculate the quadrature sum of $h_\mathrm{shot} (f)$ in both arms of the interferometer. The shot noise can be regarded as the statistical fluctuations of the photon number at the photodetector. The minimum phase change $\phi_\mathrm{shot}$ when the laser light is detected at the photodetector \cite{Takaki} is given by

\begin{equation}
    \delta \phi_\mathrm{shot} = \sqrt{\frac{\hbar \Omega}{2 \eta P}}
  \label{eq:32} \mathrm{,}
\end{equation}

\noindent quantum efficiency of the photodetector is $\eta$, and the laser power at the photodetector is $P_\mathrm{PD}$. The angular frequency of the laser, $\Omega$, is given by

\begin{equation}
    \Omega = 2 \pi \frac{c}{\lambda}
  \label{eq:33} \mathrm{.}
\end{equation}

Now we calculate $h_\mathrm{shot} (f)$, which is equivalent to $\phi_\mathrm{shot}$. Assuming that the phase of the light in the interferometer shifts by $\phi(t)$ when the gravitational wave reaches the interferometer, the electric field $E_\mathrm{PD}$ is given by Eq.(31). The phase shift $\phi(t)$ is then given by

\begin{equation}
    \phi(t) = \frac{1}{-r_1 + \alpha_\mathrm{C}} \int_{-\infty}^{\infty} H_\mathrm{FP} (\omega) h (\omega) \mathrm{e}^{i \omega t} \mathrm{d} \omega
  \label{eq:34} \mathrm{,}
\end{equation}

where $H_\mathrm{FP} (\omega)$, the transfer function between the strain and the phase, is given by \cite{Takaki}

\begin{equation}
    H_\mathrm{FP} (\omega) = \frac{2 \alpha_\mathrm{C} \Omega}{\omega} \cdot \frac{\sin\left( \frac{L \omega}{c} \right) \mathrm{e}^{-i \frac{L \omega}{c}} }{1 - r_{\mathrm{eff},1} r_{\mathrm{eff},2} \mathrm{e}^{-i \frac{2L \omega}{c}}}
  \label{eq:35} \mathrm{.}
\end{equation}

With the coefficient, $\mathrm{F}$, given by Eq.(16), the absolute value of $H_\mathrm{FP} (\omega)$ is given by

\begin{equation}
    \left|H_\mathrm{FP} (\omega)\right| = \frac{2 \alpha_\mathrm{C} \Omega}{\omega \left( 1 - r_{\mathrm{eff},1} r_{\mathrm{eff},2} \right)} \cdot \frac{\left| \sin \left( \frac{L \omega}{c} \right) \right|}{\sqrt{1 + \mathrm{F} \sin^2 \left( \frac{L \omega}{c} \right)}}
  \label{eq:36} \mathrm{.}
\end{equation}

Then Eq.(36) is rewritten as Eq.(37) on the assumption of $L\omega / c \ll 1$.

\begin{equation}
    \left|H_\mathrm{FP} (f)\right| \approx \frac{4 \pi L}{\lambda} \cdot \frac{\alpha_\mathrm{C}}{1 - r_{\mathrm{eff},1} r_{\mathrm{eff},2}} \cdot \frac{1}{\sqrt{1 + \left( \frac{f}{f_\mathrm{p}} \right)^2}}
  \label{eq:37} \mathrm{,}
\end{equation}

with $f_\mathrm{P}$ defined by

\begin{equation}
    f_\mathrm{p} \equiv \frac{c}{4 \mathcal{F}_\mathrm{eff} L}
  \label{eq:38} \mathrm{.}
\end{equation}

The gravitational wave strain $h_\mathrm{shot} (f)$, which is equivalent to the phase change $\phi_\mathrm{shot} (f)$ at a certain frequency $f$, is given by \cite{Takaki}

\begin{equation}
    h_\mathrm{shot} (f) = \left| \frac{-r_1 + \alpha_\mathrm{C}}{H_\mathrm{FP} (f)} \right| \delta \phi_\mathrm{shot} (f)~.
  \label{eq:39}
\end{equation}

Using Eq.(14), $P_\mathrm{PD}$ is given by

\begin{equation}
    P_\mathrm{PD} = {D_1}^2 \left| -r_1 + \alpha_\mathrm{C} \right|^2 P_\mathrm{in}
  \label{eq:40} \mathrm{.}
\end{equation}

Using Eqs.(37)-(40), $h_\mathrm{shot} (f) $ can be written as

\begin{align}
    h_\mathrm{shot} (f) &= \frac{\lambda}{4 \pi L} \left| \frac{\left( 1 - r_{\mathrm{eff},1} r_{\mathrm{eff},2} \right) \left( -r_1 + \alpha_\mathrm{C} \right)}{\alpha_\mathrm{C}} \right| \sqrt{\frac{\hbar \Omega}{2 \eta {D_1}^2 \left( -r_1 + \alpha_\mathrm{C} \right)^2 P_\mathrm{in}}} \sqrt{1 + \left( \frac{f}{f_\mathrm{p}} \right)^2} \notag \\
        &= \frac{\sqrt{\lambda}}{4 \pi L} \left| \frac{1 - r_{\mathrm{eff},1} r_{\mathrm{eff},2}}{\alpha_\mathrm{C}} \right| \sqrt{\frac{\pi \hbar c}{\eta {D_1}^2 P_\mathrm{in}}} \sqrt{1 + \left( \frac{f}{f_\mathrm{p}} \right)^2}
  \label{eq:41} \mathrm{.}
\end{align}

Then, because the shot noise from each arm is uncorrelated, the total shot noise, $h'_\mathrm{shot} (f)$, can be written as the quadrature sum of the shot noise from each arm, which is given by

\begin{equation}
    h'_\mathrm{shot} (f) = \frac{\sqrt{\lambda}}{4 \pi L} \frac{\left( 1 - r_{\mathrm{eff},1} r_{\mathrm{eff},2} \right)^2}{(t_1 D_1)  {t_{\mathrm{eff},1}} r_{\mathrm{eff},2}} \sqrt{\frac{4 \pi \hbar c}{\eta P_0}} \sqrt{1 + \left( \frac{f}{f_\mathrm{p}} \right)^2}~.
  \label{eq:42}
\end{equation}

Here $P_\mathrm{in}$ is converted to $P_0$, which is the total laser power of FPMI. The pre-conceptual design of DECIGO is shown as a reference in Fig. 4. DECIGO uses a differential FP interferometer, whose quantum noise is in principle the same as that of the FPMI.

\begin{figure}[htbp]
  \centering
   \includegraphics[clip,width=8cm]{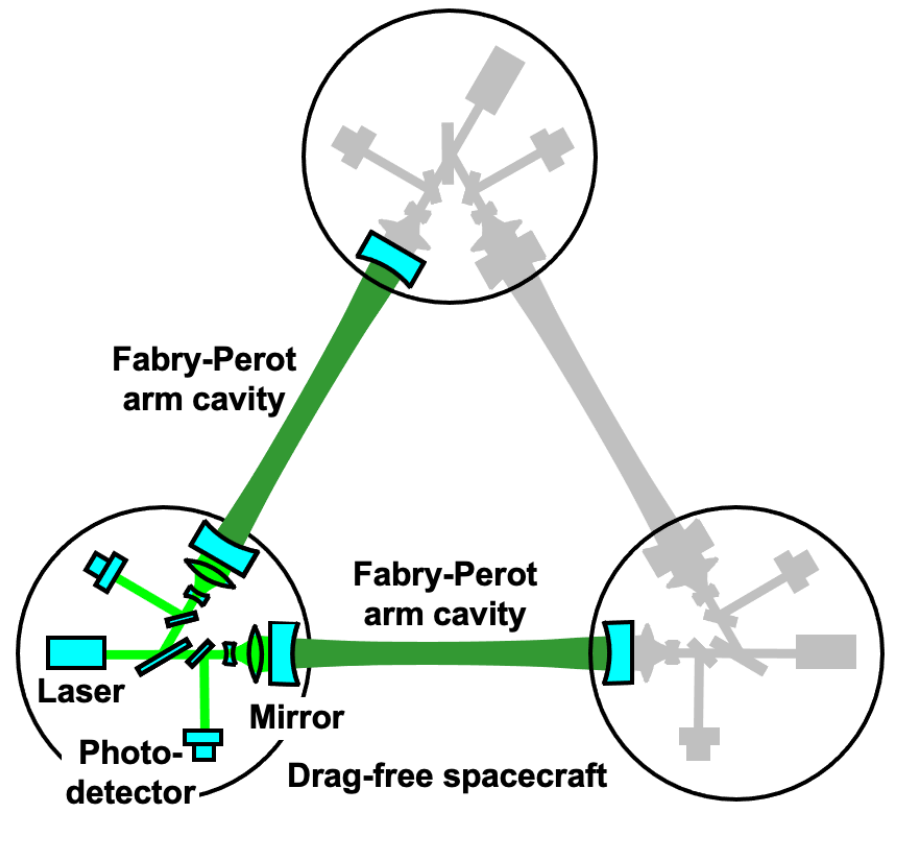}
   \label{fig:4}
   \caption{Differential FP interferometer used in the pre-conceptual design of DECIGO. The laser beam is divided into two beams by the beam splitter. Each arm has each photodetector and FP cavity.}
 \end{figure}

The radiation pressure noise of a FP cavity is derived from fluctuations of the mirror positions due to fluctuations of the laser power. The fluctuations of the laser power can be attributed to statistical fluctuations of the number of photons. The mirror is subject to the force from the laser radiation pressure. When the laser power is $P$, its force is represented as $2P/c$. The position $x$ of the mirror follows the equation of motion, which is represented as

\begin{equation}
    M \frac{\mathrm{d}^2 x}{\mathrm{d} t^2} = \frac{2 P}{c}
  \label{eq:43} \mathrm{.}
\end{equation}

The relationship between the fluctuation of the mirror position $\delta x(\omega)$ and the laser power, which is derived from the Fourier expansion of Eq.(43), is given by

\begin{equation}
    \delta x (\omega)= \frac{2}{M \omega^2 c} \delta P(\omega)~~~~(\omega \geqq 0)
  \label{eq:44} \mathrm{.}
\end{equation}

The energy per photon is $\hbar \Omega$, so the laser power $P$ can be written in terms of the number of photons, $N$, as

\begin{equation}
    P = N \hbar \Omega = N \frac{2\pi \hbar c}{\lambda}~~~~(-\infty < \omega < \infty)
  \label{eq:45} \mathrm{.}
\end{equation}

The fluctuation of $N$ is proportional to the square root of $N$, that is

\begin{equation}
    N = \frac{2 P}{\hbar \Omega} \pm \sqrt{\frac{2 P}{\hbar \Omega}} = \frac{P \lambda}{\pi \hbar c} \pm \sqrt{\frac{P  \lambda}{\pi \hbar c}}~~~~(\omega \geqq 0)
  \label{eq:46} \mathrm{.}
\end{equation}

Then the power fluctuation $\delta P(\omega)$ is given by

\begin{equation}
    \delta P(\omega) = \sqrt{\frac{P \lambda}{\pi \hbar c}} \cdot \frac{2\pi \hbar c}{\lambda} = \sqrt{\frac{4 \pi \hbar c P}{\lambda}}~.
  \label{eq:47}
\end{equation}

Using Eqs.(44) and (47), the fluctuation of the mirror position $\delta P(\omega)$ is given by

\begin{equation}
    \delta x (\omega) = \frac{2}{M \omega^2 c} \sqrt{\frac{4\pi \hbar c P}{\lambda}} = \frac{4}{M \omega^2 } \sqrt{\frac{\pi \hbar P}{c \lambda}}~.
  \label{eq:48}
\end{equation}

In terms of the FP cavity, whose arm length is $L$, the response from the gravitational wave with the amplitude of $\delta x / L$ is equal to the one from $\delta x$. $\delta x$ is the fluctuation of the mirror position in the FP cavity \cite{Takaki}. For this reason, $h_\mathrm{rad} (f)$, which corresponds to the phase change by $\delta x$, is represented as

\begin{equation}
    h_\mathrm{rad} (f) = \frac{\delta x}{L} = \frac{4}{L M (2\pi f)^2 }\sqrt{\frac{\pi \hbar P}{c \lambda}}
  \label{eq:49} \mathrm{.}
\end{equation}

$P$ in Eq.(49) has contributions from two sources: the light reflected at the input mirror and the laser light circulating inside the FP cavity. As a result, the total radiation pressure noise of an arm cavity in FPMI is derived from two sources. The laser power reflected at the input mirror is negligible because this power is much less than the laser power inside the FP cavity under the condition that the FP cavity has relatively high finesse. The laser power reflected at the end mirror is defined as $P_\mathrm{E}$, and that at the input mirror is defined as $P_\mathrm{F}$. Using Eq.(10), electric fields, $E_\mathrm{E}$ and $E_\mathrm{F}$, are given by

\begin{align}
    E_\mathrm{E} &= r_2 D_2 \mathrm{e}^{i \frac{\phi}{2}} \cdot E_1 \notag \\
      &= \frac{t_{\mathrm{eff},1} (r_2 D_2) \mathrm{e}^{i \frac{\phi}{2}}}{1 - r_{\mathrm{eff},1} r_{\mathrm{eff},2} \cdot \mathrm{e}^{i \phi}} E_\mathrm{in}~
  \label{eq:50} \mathrm{and}
\end{align}

\begin{align}
    E_\mathrm{F} &= r_{\mathrm{eff},2} (r_1 D_1) \mathrm{e}^{i \phi} \cdot E_1 \notag \\
      &= \frac{t_{\mathrm{eff},1} r_{\mathrm{eff},2} (r_1 D_1) \mathrm{e}^{i \phi}}{1 - r_{\mathrm{eff},1} r_{\mathrm{eff},2} \cdot \mathrm{e}^{i \phi}} E_\mathrm{in}~
  \label{eq:51} \mathrm{.}
\end{align}

In Eq.(50) and Eq.(51), we treat only the leakage loss as the diffraction loss because the radiation pressure noise is caused by the laser power, which is just after the reflection. For this reason, $E_\mathrm{E}$ is $E_1$ multiplied by the reflectivity $r_2$ of the end mirror and the coefficient $D_i$ of the leakage loss. Also, $E_\mathrm{F}$ is $E_1$ multiplied by the effective reflectivity $r_{\mathrm{eff,2}}$ of the end mirror, the reflectivity $r_1$ of the input mirror, and the coefficient $D_i$ of the leakage loss. Using Eqs.(50)-(51), $P_\mathrm{E}$ and $P_\mathrm{F}$ can be written as

\begin{equation}
    P_\mathrm{E} = \frac{{t_{\mathrm{eff},1}}^2 (r_2 D_2)^2}{(1-r_{\mathrm{eff},1} r_{\mathrm{eff},2})^2 \left[ 1 + \mathrm{F}\sin^2 \left( \frac{\phi}{2} \right) \right]} P_\mathrm{in}
  \label{eq:52} \  \mathrm{and}
\end{equation}

\begin{equation}
    P_\mathrm{F} = \frac{{t_{\mathrm{eff},1}}^2 {r_{\mathrm{eff},2}}^2 (r_1 D_1)^2}{(1-r_{\mathrm{eff},1} r_{\mathrm{eff},2})^2 \left[ 1 + \mathrm{F}\sin^2 \left( \frac{\phi}{2} \right) \right]} P_\mathrm{in}
  \label{eq:53} \mathrm{.}
\end{equation}

Here we define $k_\mathrm{E}$ and $k_\mathrm{F}$, which is given by

\begin{equation}
    k_\mathrm{E} \equiv \frac{P_\mathrm{in}}{1 + \mathrm{F} \sin^2 \left( \frac{\phi}{2} \right)}~
  \label{eq:54} \ \mathrm{and}
\end{equation}

\begin{equation}
    k_\mathrm{F}  \equiv \frac{P_\mathrm{in}}{1 + \mathrm{F} \sin^2 \left( \frac{\phi}{2} \right)}~
  \label{eq:55} \mathrm{.}
\end{equation}

With Eqs.(54)-(55), Eqs.(52)-(53) can be written as

\begin{equation}
    P_\mathrm{E} = k_\mathrm{E} \frac{P_\mathrm{in}}{1 + \mathrm{F} \sin^2 \left( \frac{\phi}{2} \right)}~
  \label{eq:56} \  \mathrm{and}
\end{equation}

\begin{equation}
    P_\mathrm{F} = k_\mathrm{F} \frac{P_\mathrm{in}}{1 + \mathrm{F} \sin^2 \left( \frac{\phi}{2} \right)}~
  \label{eq:57} \mathrm{.}
\end{equation}

Substituting $P_\mathrm{E}$ and $P_\mathrm{F}$ into Eq.(47), the fluctuation of each conponent of laser power is given by

\begin{equation}
    \delta P_\mathrm{E} = k_\mathrm{E} \sqrt{\frac{P_\mathrm{in}}{1 + \mathrm{F} \sin^2 \left( \frac{\phi}{2} \right)}}
  \label{eq:58} \  \mathrm{and}
\end{equation}

\begin{equation}
    \delta P_\mathrm{F} = k_\mathrm{F} \sqrt{\frac{P_\mathrm{in}}{1 + \mathrm{F} \sin^2 \left( \frac{\phi}{2} \right)}}
  \label{eq:59} \mathrm{.}
\end{equation}

The term in the square root represents the noise caused by a single reflection. This term is multiplied by the terms related to the finesse, $k_\mathrm{E}$ and $k_\mathrm{F}$, to represent the fluctuation of the laser power inside the FP cavity. Thus, the radiation pressure noise $h_\mathrm{rad} (f)$ of one arm FP cavity is given by

\begin{align}
    h_\mathrm{rad} (f) &= \frac{2}{L M c (2\pi f)^2 } \delta P_\mathrm{E} + \frac{2}{L M c (2\pi f)^2} \delta P_\mathrm{F} \notag \\
      &= \frac{4}{L M (2\pi f)^2} \sqrt{\frac{\pi \hbar}{c \lambda}} \left( k_\mathrm{E} + k_\mathrm{F} \right) \sqrt{\frac{P_\mathrm{in}}{1 + \mathrm{F} \sin^2 \left( \frac{\phi}{2} \right)}} \notag \\
      &= \frac{4}{L M (2\pi f)^2} \cdot \frac{{t_{\mathrm{eff},1}}^2 \cdot (r_2 D_2)^2 \cdot \left( 1 + (r_1 D_1 D_2)^2 \right)}{\left(1 - r_{\mathrm{eff},1} r_{\mathrm{eff},2} \right)^2} \sqrt{\frac{\pi \hbar P_\mathrm{in}}{c \lambda}} \frac{1}{\sqrt{1 + \mathrm{F}\sin^2(\frac{\phi}{2})}}~.
  \label{eq:60}
\end{align}

When $\phi =2L \Omega /c$ is substituted into Eq.(58), assuming that $L\omega / c \ll 1$, it can be rewritten as

\begin{equation}
    h_\mathrm{rad} (f) \approx \frac{8}{L M (2\pi f)^2} \cdot \frac{{t_{\mathrm{eff},1}}^2 \cdot (r_2 D_2)^2 \cdot \left( 1 + (r_1 D_1 D_2)^2 \right)}{\left(1 - r_{\mathrm{eff},1} r_{\mathrm{eff},2} \right)^2} \sqrt{\frac{\pi \hbar P_\mathrm{in}}{c \lambda}} \frac{1}{\sqrt{1 + \left( \frac{f}{f_\mathrm{p}} \right)^2}}
  \label{eq:61} \mathrm{.}
\end{equation}

Finally, the total radiation pressure noise in a FPMI, $h'_\mathrm{rad} (f)$, with no correlation between the noises in the two arms is given by

\begin{align}
    h'_\mathrm{rad} (f) &= \sqrt{2} \frac{4}{L M (2\pi f)^2} \cdot \frac{{t_{\mathrm{eff},1}}^2 \cdot (r_2 D_2)^2 \cdot \left( 1 + (r_1 D_1 D_2)^2 \right)}{\left(1 - r_{\mathrm{eff},1} r_{\mathrm{eff},2} \right)^2} \sqrt{\frac{\pi \hbar \frac{P_0}{2}}{c \lambda}} \frac{1}{\sqrt{1 + \left( \frac{f}{f_\mathrm{p}} \right)^2}} \notag \\
      &= \frac{4}{L M (2\pi f)^2} \cdot \frac{{t_{\mathrm{eff},1}}^2 \cdot (r_2 D_2)^2 \cdot \left( 1 + (r_1 D_1 D_2)^2 \right)}{\left(1 - r_{\mathrm{eff},1} r_{\mathrm{eff},2} \right)^2} \sqrt{\frac{\pi \hbar P_0}{c \lambda}} \frac{1}{\sqrt{1 + \left( \frac{f}{f_\mathrm{p}} \right)^2}}
  \label{eq:62} \mathrm{.}
\end{align}

Assuming that the diffraction is negligible, and the reflectivity $r_2$ is equal to $1$, Eqs.(42) and (62) can be written as the calculation results, $h''_\mathrm{shot} (f)$ and $h''_\mathrm{rad} (f)$, which are written by

\begin{equation}
    h''_\mathrm{shot} (f) = \frac{\sqrt{\lambda}}{4 \pi L} \frac{(1-r_1)^2}{{t_1}^2} \sqrt{\frac{4 \pi \hbar c}{\eta P_0}} \sqrt{1 + \left( \frac{f}{f_\mathrm{p}} \right)^2}~
  \label{eq:63}  \mathrm{and}
\end{equation}

\begin{align}
    h''_\mathrm{rad} (f) &= \frac{4}{L M (2 \pi f)^2} \cdot \frac{{t_1}^2 \left( 1 + {r_1}^2 \right)}{(1 - r_1)^2} \sqrt{\frac{\pi \hbar P_0}{c \lambda}} \frac{1}{\sqrt{1 + \left( \frac{f}{f_\mathrm{p}} \right)^2}}
  \label{eq:64}  \mathrm{.}
\end{align}

Assuming that $r_1 \approx 1$,

\begin{equation}
    \frac{{t_1}^2 r_2}{(1-r_1 r_2)^2} \approx \frac{2 \mathcal{F}}{\pi}
  \label{eq:65} \mathrm{,}
\end{equation}

and Eq.(63) and Eq.(64) are rewritten as

\begin{equation}
    h''_\mathrm{shot} (f) \approx \frac{1}{4 \mathcal{F} L} \sqrt{\frac{\pi \hbar c \lambda}{\eta P_0}} \sqrt{1 + \left( \frac{f}{f_\mathrm{p}} \right)^2}
  \label{eq:66}  \mathrm{and}
\end{equation}

\begin{equation}
    h''_\mathrm{rad} (f) \approx \frac{16 \mathcal{F}}{L M (2 \pi f)^2} \sqrt{\frac{\hbar P_0}{\pi c \lambda}} \frac{1}{\sqrt{1 + \left( \frac{f}{f_\mathrm{p}} \right)^2}}
  \label{eq:67}  \mathrm{.}
\end{equation}

These calculation results are consistent with \cite{Kimble}.

\section{Quantum noise in DECIGO}

\ \ \ \ We now use the default parameters of DECIGO to calculate the quantum noise of DECIGO. First, we calculate the power spectral density (PSD) of the quantum noise using Eq.(42) and Eq.(62). The PSD is given by

\begin{equation}
  S_\mathrm{h} (f) = h'_\mathrm{shot} (f)^2 + h'_\mathrm{rad} (f)^2
     \label{eq:66}  \mathrm{.}
\end{equation}

We define two noise PSDs, $S_\mathrm{h}$ and $S_\mathrm{h,eff}$, for comparison between the quantum noise without and with the diffraction. In Fig. 5, we plot the noise spectra in the FPMI for $ \sqrt{S_\mathrm{h}}$ and $\sqrt{S_\mathrm{h,eff}}$. The parameters used by calculation are shown in Table 1. \\
\ \ \ \ Figure 5 shows two curves: the black one shows the noise spectra with no diffraction, the magenta one shows the noise spectra with diffraction. The diffraction causes the reduction of the laser power. At the frequencies between $10^{-3}$ and $10^{-1}$ Hz, the magenta curve is lower than the black one because the effective laser power is smaller with diffraction. For the same reason, at frequencies above $10^{-1}$ Hz, the magenta curve is higher than the black one.

\begin{table}[h]
  \centering
    \begin{tabular}{|c|c|c|}
      symbol & Default(w/o diffraction) & Default(w/ diffraction) \\ \hline
      $L$ & 1000km & 1000km \\ \hline
      $r$ & 0.855 & 0.855 \\ \hline
      $D$ & 1 & 0.9760 \\ \hline
      $P$ & 10W & 10W \\ \hline
      $\mathcal{F}$ & 10 & 7.611 \\
    \end{tabular}
    \caption{Mechanical and optical default parameters of DECIGO. $L$ is the cavity length, $r$ is the reflectivity of the mirror, $D$ is the coefficient of the diffraction, $P$ is the laser power, and $\mathcal{F}$ is the finesse of the FP cavity. The finesse is calculated in the two cases: with and without the diffraction loss.}
\end{table}

\begin{figure}[h]
  \centering
   \includegraphics[clip,width=10cm]{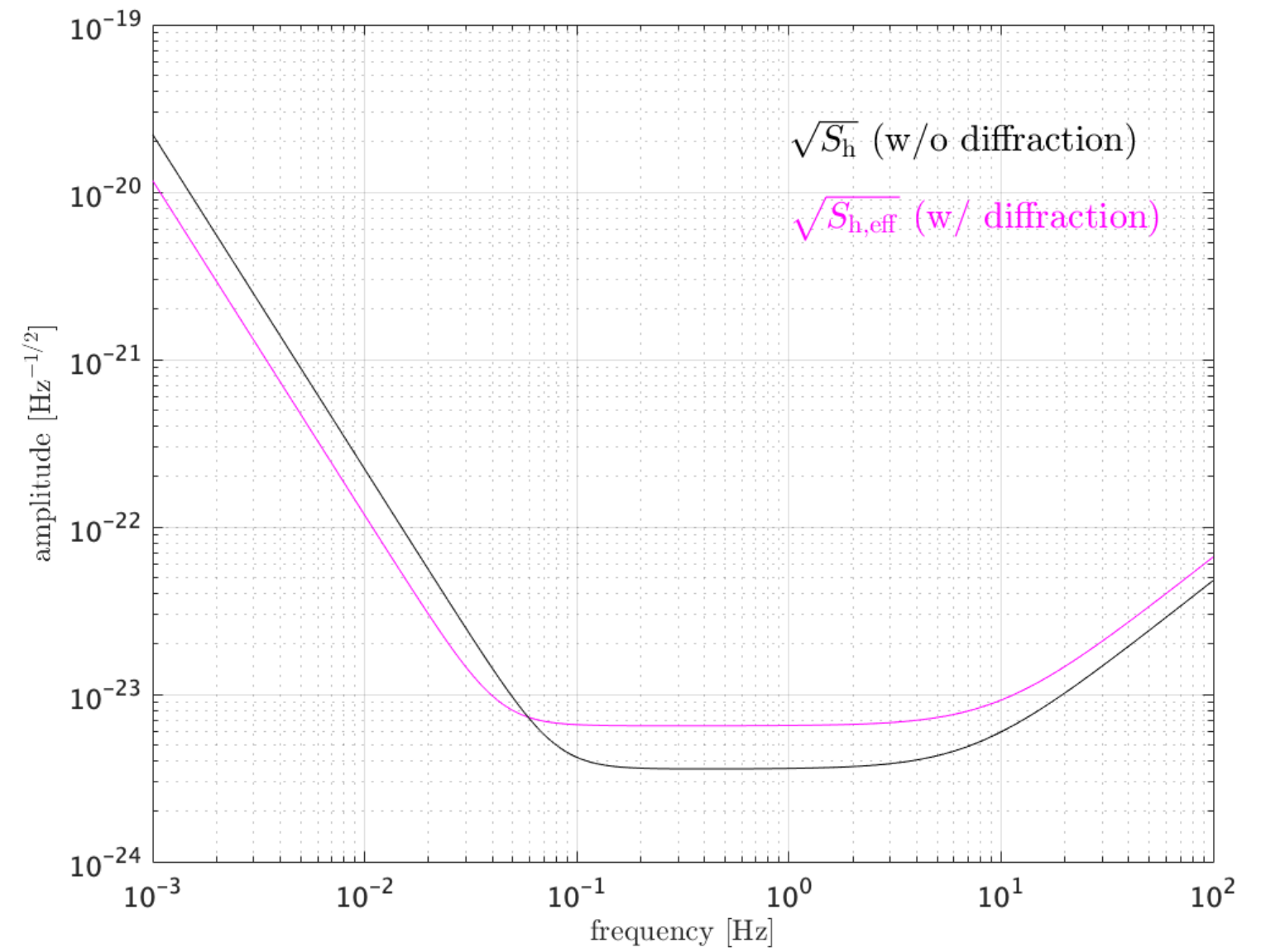}
   \label{fig:5}
   \caption{Sensitivity curves in terms of the square root of the PSD without the diffraction loss (black) and the one including the effect of the diffraction (magenta). Each parameter in the calculation is shown in Table. 1.}
 \end{figure}



\section{Summary}

\ \ \ \ In this paper, the treatment of diffraction loss in a FP cavity and quantum noise, including the effect of the diffraction loss, are presented. First, two kinds of diffraction losses are treated: leakage loss and higher-order mode loss. The coefficient of the diffraction loss is defined as $D_i$, which is given by Eq.(7). The reflectivity and transmissivity influenced by the diffraction loss are defined as the effective reflectivity $r_\mathrm{eff}$ and the effective transmissivity $t_\mathrm{eff}$ for each mirror in the cavity. $r_\mathrm{eff}$ and $t_\mathrm{eff}$ are given by Eq.(8) and Eq.(9). Also the finesse influenced by the diffraction loss is defined as $\mathcal{F}_\mathrm{eff}$, which is given by Eq.(23). In terms of Eqs.(8), (9), and (23), a FP cavity with diffraction loss can be treated with the coefficient $D_\mathrm{i}$. Also, quantum noise, including the effect of diffraction loss, can be treated with Eqs.(7)-(9) and (23). The shot noise and the radiation pressure noise including the diffraction loss given by Eq.(42) and Eq.(62). This result is useful for optimization of design of DECIGO optical  parameters \cite{Ishikawa}. This method is also applicable to all FP cavities with a relatively high finesse and a significant diffraction loss in any interferometer.

\section*{Acknowledgement}

We would like to thank Rick Savage for English editing. We would like to thank Naoki Seto for helpful discussion. This work was supported by the Japan Society for the Promotion of Science (JSPS) KAKENHI Grant Number JP19H01924.

\end{document}